\documentclass[preprint,twocolumn,sort&compress,3p,times]{elsarticle}

\usepackage[hyperindex,breaklinks]{hyperref}
\usepackage{atlasphysics}

\usepackage{amsmath,graphicx}
\usepackage{a4wide}

\usepackage{booktabs}
\usepackage{numcompress}
\usepackage{multirow,xspace}
\usepackage{rotating}
\usepackage{subfigure}

\graphicspath{{./figures/}}

\newcommand{\metrel}{\ensuremath{E_{\rm T,rel}^{\rm miss}}}
\newcommand{\mT}{\ensuremath{m_{\rm T}}}
\newcommand{\pTll}{\ensuremath{p_{\rm T}^{\ell\ell}}}
\newcommand{\ptll}{\pTll}
\newcommand{\vpTll}{\ensuremath{{\bf p}_{\rm T}^{\ell\ell}}}

\newcommand{\mll}{\ensuremath{m_{\ell\ell}}\xspace}

\newcommand{\hwwlnln}{\ensuremath{H{\rightarrow\,}WW^{(\star)}{\rightarrow\,}\ell\nu\ell\nu}}

\newcommand{\ZeroJet}{\mbox{0-jet}}
\newcommand{\OneJet}{\mbox{1-jet}}
\newcommand{\TwoJet}{\mbox{2-jet}}

\newcommand{\mhsplit}{200}

\newcommand{\limobslow}{133}
\newcommand{\limobshigh}{261}
\newcommand{\limexplow}{127}
\newcommand{\limexphigh}{233}

\usepackage{preprintcover}  % load package
\PreprintCoverPaperTitle{Search for the Standard Model Higgs boson in 
  the $\boldsymbol{H\to WW^{(\star)}\to \ell\nu\ell\nu}$ decay mode with
  4.7~fb$^{\boldsymbol{-1}}$ of ATLAS data \\ at $\boldsymbol{\sqrt{s}=7 }$ \TeV{} }
\PreprintIdNumber{CERN-PH-EP-2012-126}  % CERN preprint number
\PreprintCoverAbstract{
  A search for the Standard Model Higgs boson in the \hwwlnln\ ($\ell=e,\mu$)
  decay mode is presented. The search is performed using proton-proton collision data corresponding 
  to an integrated luminosity of 4.7~$\ifb$ at a centre-of-mass energy of 7~\TeV{}
  collected during 2011 with the ATLAS detector at the Large Hadron Collider.
  No significant excess of events over the expected background is observed.
  An upper bound is placed on the Higgs boson production cross section as a
  function of its mass.
  A Standard Model Higgs boson with mass in the range between \limobslow~\GeV{} and \limobshigh~\GeV{} is
  excluded at 95\% confidence level, while the expected
  exclusion range is from \limexplow~\GeV{} to \limexphigh~\GeV.}
\PreprintJournalName{Physics Letters B}
\journal{Physics Letters B}

\title{\textbf{Search for the Standard Model Higgs boson in
  the $\boldsymbol{H\to WW^{(\star)}\to \ell\nu\ell\nu}$ decay mode with
 4.7~fb$^{\boldsymbol{-1}}$  of ATLAS data \\ at
 $\boldsymbol{\sqrt{s}=7 }$ \TeV{}}}
\author{The ATLAS Collaboration}

\begin{document}

%%%%%%%%%%%%%%%%%%%%%%%%%%%%%%%%%%%%
%           Title page             % 
%%%%%%%%%%%%%%%%%%%%%%%%%%%%%%%%%%%%

%\skipbeforetitle{100pt}

\begin{frontmatter}  % for elsarticle

% Title
\title{\textbf{Search for the Standard Model Higgs boson in
  the $\boldsymbol{H\to WW^{(\star)}\to \ell\nu\ell\nu}$ decay mode with
 4.7~fb$^{\boldsymbol{-1}}$  of ATLAS data at
 $\boldsymbol{\sqrt{s}=7 }$ \TeV{}}}
% Author
% \author{The ATLAS Collaboration}
% \address[aff]{CERN, Geneva, Switzerland}

% Abstract
\begin{abstract}
  A search for the Standard Model Higgs boson in the \hwwlnln\ ($\ell=e,\mu$)
  decay mode is presented. The search is performed using proton-proton collision data corresponding 
  to an integrated luminosity of $4.7\,\ifb$ at a centre-of-mass energy of 7~\TeV{}
  collected during 2011 with the ATLAS detector at the Large Hadron Collider.
  No significant excess of events over the expected background is observed.
  An upper bound is placed on the Higgs boson production cross section as a
  function of its mass.
  A Standard Model Higgs boson with mass in the range between $\limobslow\GeV$ and $\limobshigh\GeV$ is
  excluded at 95\% confidence level, while the expected
  exclusion range is from $\limexplow\,\GeV$ to $\limexphigh\,\GeV$.
\end{abstract}

\begin{keyword}
  ATLAS \sep LHC \sep Higgs \sep WW
  \PACS{14.80.Bn, 12.15.Ji, 14.70.Fm}
\end{keyword}

\end{frontmatter}  % for elsarticle

%%%%%%%%%%%%%%%%%%%%%%%%%%%%%%%%%%%%
%            Content               % 
%%%%%%%%%%%%%%%%%%%%%%%%%%%%%%%%%%%%

\section{Introduction}
\label{sec:intro}

The Higgs boson is the only elementary particle in the Standard Model (SM) of particle
physics that has not yet been observed.
It is intimately related to the Higgs
mechanism~\cite{prl_13_321,prl_13_508,Guralnik:1964eu}
which in the SM gives mass to all other massive elementary particles.  The search for
this particle is a centrepiece of the Large Hadron Collider (LHC) physics programme.

Indirect limits on the Higgs boson mass of 
$m_H < 158\GeV$ at $95\%$ confidence level (CL) have been set using global
fits to precision electroweak results~\cite{lepew:2010vi}.
Direct searches at LEP and the Tevatron have excluded at 95\% 
CL a SM Higgs boson with a mass below $114.4\GeV$~\cite{Barate:2003sz}
and in the regions $147\GeV < m_H < 179\GeV$ and
$100\GeV < m_{H} < 106\GeV$~\cite{TheTEVNPHWorkingGroup:2012nc}, respectively.

The results of searches in various channels using data corresponding to an
integrated luminosity of approximately $5\,\ifb$ have been
reported recently by
the ATLAS Collaboration, excluding the mass ranges 112.9~\GeV--115.5~\GeV,
131~\GeV--238~\GeV, and 251~\GeV--466~\GeV~\cite{ATLAS-f5bHiggsCombination}; and 
by the CMS Collaboration, excluding the mass range from 127~\GeV{} to
600~\GeV~\cite{CMS-f5bHiggsCombination}.

In the $\hwwlnln$ channel (with $\ell=e,\,\mu$), ATLAS reported the results of a
search using the first $2.05\,\ifb$ of data from 2011, which excluded a SM Higgs boson in the mass
range $145\GeV < m_{H} < 206\GeV$ at 95\% CL~\cite{ATLAS-2fbHWW}.
The analysis described in this Letter uses the full 2011 dataset, which after
requiring that all detector components are fully functional corresponds to
$4.7\,\ifb$ of proton-proton ($pp$) collisions at $\sqrt{s}=7\TeV$.
The selection criteria described in  Ref.~\cite{ATLAS-2fbHWW} are modified to gain sensitivity at low $m_{H}$ and to
cope with increased instantaneous luminosities.  The previous cut-based approach
is extended by adding events with two jets
and by fitting for the presence of a signal using a transverse mass variable.
A similar search has been performed by the CMS Collaboration~\cite{Chatrchyan:2012ty}.

\section{Data and Simulated Samples}
\label{sec:samples}

The data used for this analysis were collected in 2011 using the ATLAS
detector, a multi-purpose particle physics experiment with a forward-backward
symmetric cylindrical geometry and near 4$\pi$ coverage in solid
angle~\cite{atlas-det}.
It consists of an inner tracking system surrounded by a thin
superconducting solenoid, electromagnetic and hadronic calorimeters, and an
external muon spectrometer incorporating large superconducting air-core
toroid magnets.
The combination of these systems provides charged particle measurements together with highly 
efficient and precise lepton measurements over the pseudorapidity\footnote{
  ATLAS uses a right-handed coordinate system with
  its origin at the nominal interaction point (IP) in the centre of the detector,
  and the $z$-axis along the beam line.
  The $x$-axis points from the IP to the centre of the LHC ring, and the
  $y$-axis points upwards.
  Cylindrical coordinates $(r,\phi)$ are used in the transverse plane, $\phi$
  being the azimuthal angle around the beam line.
  The pseudorapidity is defined in terms of the polar angle $\theta$ as
  $\eta=-\ln\tan(\theta/2)$.
} range $|\eta| < 2.5$.
Jets are reconstructed over the full coverage of the calorimeters, $|\eta| < 4.9$;
this calorimeter coverage also provides a precise measurement of the missing transverse momentum. 

The data used in the present analysis were collected using inclusive single-muon and single-electron triggers.
The single-muon trigger required the transverse momentum of the muon with respect
to the beam line, \pt, to exceed 18~\GeV; for the single-electron trigger the 
threshold varied from 20 to 22~\GeV. The trigger object quality requirements were
tightened throughout the data-taking period to cope with the increasing instantaneous
luminosity.

In this analysis, the signal contributions that are considered include the dominant gluon fusion production process
($gg\to H$, denoted as ggF), the vector-boson fusion production process
($qq'\to qq'H$, denoted as VBF) and the  Higgs-strahlung process ($qq'\to WH, ZH$, denoted as
$WH$/$ZH$).
For the decay of the Higgs boson, only the $\hwwlnln$ mode
is considered, with final states featuring two charged leptons ($\ell =
e,\,\mu$, including small contributions from leptonic $\tau$ decays). 
The branching fraction for this decay, as a function of $m_{H}$, is taken from
the HDECAY~\cite{Djouadi:1997yw} program.

The signal cross section is computed  to 
next-to-next-to-leading order (NNLO)~\cite{Djouadi:1991tka,Dawson:1990zj,Spira:1995rr,Harlander:2002wh,Anastasiou:2002yz,Ravindran:2003um} 
in QCD for the ggF process. Next-to-leading order (NLO) electroweak (EW) corrections are also
applied~\cite{Aglietti:2004nj,Actis:2008ug}, as well as QCD soft-gluon resummations up to 
next-to-next-to-leading log (NNLL)~\cite{Catani:2003zt}. These calculations are detailed in 
Refs.~\cite{Anastasiou:2008tj,deFlorian:2009hc,Baglio:2010ae}, and assume factorisation between QCD and EW corrections.
Full NLO QCD and EW corrections~\cite{Ciccolini:2007jr,Ciccolini:2007ec,Arnold:2008rz} and approximate NNLO QCD corrections
~\cite{Bolzoni:2010xr} are used to calculate the cross sections for VBF signal production.
The cross sections of the associated $WH/ZH$ production processes are calculated
up to NNLO QCD corrections~\cite{Han:1991ia,Brein:2003wg} and NLO EW corrections~\cite{Ciccolini:2003jy}.

The Monte Carlo (MC) generators used to model signal and background processes
are listed in Table~\ref{tab:cross}. For most processes, separate programs are used to generate 
the hard scattering process and to model the parton showering and
hadronisation stages. Wherever HERWIG~\cite{herwig} is used for the latter,
JIMMY~\cite{jimmy} is used for the simulation of the underlying event. The MLM
matching scheme~\cite{alwall-2008-53} is used for the description of the
$W$+jets and $Z/\gamma^{\star}$+jets processes.
\begin{table}[htb]
  \centering
 \caption{MC generators used to model the signal and background
    processes, and corresponding cross sections (given for both $m_{H}=125\GeV$
    and $m_{H}=240\GeV$ in the case of the signal processes).
    The ggF Higgs boson \pt{} spectrum is reweighted to agree with the 
    prediction from HqT~\cite{deFlorian:2011xf}.
    All three single-top production channels ($s$-channel, $t$-channel, and
    $Wt$) are included. The number quoted for the inclusive $Z/\gamma^{\star}$ process
    (also referred to in the text as the Drell-Yan process) is for generated
    dilepton invariant masses exceeding 10~\GeV. Kinematic criteria are also applied
    in the generation of $W(\rightarrow\ell\nu)\gamma$ events (the photon must have
    $\pt > 10\GeV$ and be separated from the charged lepton by
    $\Delta R = \sqrt{(\Delta\eta^{2})+(\Delta\phi^{2})} > 0.1$) and
    $W(\rightarrow\ell\nu)\gamma^{\star}(\rightarrow\ell^{\prime}\ell^{\prime})$ events
    (the higher and lower transverse momenta of the leptons from the $\gamma^{\star}$
    decay must exceed $15\GeV$ and $5\GeV$, respectively).
    Leptonic decay modes (charged leptonic decay modes only for
    $Z/\gamma^{\star}$ production) are summed over, except for $\ttbar$, single-top, $WZ$,
    and $ZZ$ production; in these cases inclusive cross sections are quoted.
    The quoted signal production cross sections include the $\hwwlnln$ branching
    fractions but no branching fractions for the $W$ and $Z$ boson in $WH/ZH$
    production.
  }
  \scalebox{0.75}{
  \begin{tabular}{llrr}
    \hline\hline
    Process & Generator & $m_{H}$ (\GeV) & \hspace*{-10mm}$\sigma\cdot\mathrm{Br}$ (pb) \\
    \hline
    \multirow{2}{*}{ggF} & POWHEG~\cite{Alioli:2008tz,Nason:2009ai}+ & \hfill 125 & 0.347\\
    & PYTHIA~\cite{pythia} & \hfill 240 &  0.265 \\
    \multirow{2}{*}{VBF} & POWHEG+ & \hfill 125& $27\cdot 10^{-3}$\\\
    & PYTHIA & \hfill 240&  $34\cdot 10^{-3}$\\
    \multirow{2}{*}{$WH/ZH$} & \multirow{2}{*}{PYTHIA} & \hfill 125 & $20\cdot 10^{-3}$\\
    & & \hfill 240 & $6\cdot 10^{-3}$\\
    \hline
    $q\bar{q}/g\rightarrow WW$& \multicolumn{2}{l}{MC@NLO~\cite{mcatnlo}+HERWIG} & 4.68 \\
    $gg\rightarrow WW$        & \multicolumn{2}{l}{GG2WW~\cite{gg2WW}+HERWIG} & 0.14 \\
    $\ttbar$                  & \multicolumn{2}{l}{MC@NLO+HERWIG} & 167 \\
    % single top NB: sum s-ch, t-ch, Wt
    $tW/tb/tqb$               & \multicolumn{2}{l}{AcerMC~\cite{Kersevan:2004yg}+PYTHIA} & 85 \\
    % W+jets NB: BR to single lepton flavour multiplied by 3 flavours
    inclusive $W$             & \multicolumn{2}{l}{ALPGEN~\cite{alpgen}+PYTHIA} & $32\cdot 10^{3}$ \\
    % Z/gamma*+jets NB: BR to single lepton flavour multiplied by 3 flavours
    inclusive $Z/\gamma^{\star}$ & \multicolumn{2}{l}{ALPGEN~\cite{alpgen}+PYTHIA} & $15\cdot 10^{3}$ \\
    $ZZ$                      & \multicolumn{2}{l}{SHERPA~\cite{Gleisberg:2008ta}} & 5.6 \\
    $WZ$                      & \multicolumn{2}{l}{MC@NLO} & 18.0 \\
    $W\gamma$                 & \multicolumn{2}{l}{ALPGEN} & 345 \\
    $W\gamma^{\star}$~\cite{Gray:2011us} & \multicolumn{2}{l}{MadGraph~\cite{Alwall:2007st,Alwall:2011uj}} & 6.5 \\
    \hline\hline
  \end{tabular}
  } \label{tab:cross}
\end{table}

The CT10 parton distribution function (PDF) set~\cite{Lai:2010vv} is used for the
MC@NLO samples,
CTEQ6L1~\cite{cteq6} for the ALPGEN, SHERPA, and MadGraph samples,
and MRSTMCal~\cite{mrst} for the PYTHIA and AcerMC samples.
Acceptances and efficiencies are obtained from a full simulation~\cite{atlassim} of the
ATLAS detector using GEANT4~\cite{GEANT4}. This includes a realistic
treatment of the event pile-up conditions (the data are affected by the detector
response to multiple $pp$ collisions occurring in the same or in different bunch
crossings) in the 2011 data; from the
first 2.3~\ifb{} to the last 2.4~\ifb{} of data taken, the average number of $pp$ 
interactions per bunch crossing increased from 6.3 to 11.6.

\section{Event Selection}
\label{sec:selection}

Events are required to have a primary vertex 
consistent with the beam spot position, 
with at least three associated tracks with $\pt > 400\MeV$.
Overall quality criteria are applied  in order to suppress
non-collision backgrounds
such as cosmic-ray muons, beam-related
backgrounds, or noise in the calorimeters.

$\hwwlnln$ candidates (with $\ell=e,\,\mu$) are pre-selected by requiring exactly two oppositely charged
leptons with \pt{} thresholds of 25 \GeV{} and 15 \GeV{} for the
leading and sub-leading lepton, respectively.
For muons, the range $|\eta|< 2.4$ is used; for electrons, the range $|\eta|<
2.47$ is used, with the region $1.37 < |\eta|< 1.52$ (corresponding to the
boundary between barrel and end-cap calorimeters) excluded.
The selected electron candidates are reconstructed using a combination of tracking and
calorimetric information~\cite{electron-performance},
while the muon candidates are identified by matching tracks reconstructed in the
inner detector and in the muon spectrometer~\cite{muon-reco}.
At least one of the selected leptons is required to match a triggering object.
Leptons from heavy-flavour decays and jets satisfying the lepton identification criteria are 
suppressed by requiring the leptons to be isolated:
the scalar sum of the \pt{} of charged particles and of the calorimeter energy deposits within 
$\Delta R=\sqrt{\Delta\phi^{2}+\Delta\eta^{2}}=0.3$ of the lepton direction are each 
required to be less than approximately 0.15 times the lepton \pt, with slight
differences between track- and calorimeter-based criteria and between electrons
and muons.

The Drell-Yan process
leads to two same-flavour, opposite-sign 
high-$\pt$ leptons. In the $ee$ and $\mu\mu$ channels (the channels are
indicated by the charged lepton flavours), this background is suppressed by requiring
the dilepton invariant mass to be greater than $12~\GeV$, and
to differ from the $Z$-boson mass $m_{Z}$ by at least 15~\GeV.
For the $e\mu$ channel, the dilepton invariant mass is required to be greater than $10 \GeV$.

Drell-Yan events and multijet production via QCD processes are suppressed by
requiring large $\met$. The \met{} is the magnitude of
$\boldsymbol{p}_{\mathrm{T}}^{\rm miss}$, the negative vector sum of the
reconstructed objects' transverse momenta, including muons, electrons,
photons, jets, and clusters of calorimeter cells not associated with these objects.
The quantity $\metrel$ used in this analysis is defined as:
$\metrel = \met\sin\Delta\phi_{\min}$, with $\Delta\phi_{\min} \equiv
\min(\Delta\phi, \frac{\pi}{2})$. Here, $\Delta\phi$ is the angle between 
$\boldsymbol{p}_{\mathrm{T}}^{\rm miss}$ 
and the transverse momentum of the 
nearest lepton or jet with $\pt > 25\GeV$.
For the $ee$ and $\mu\mu$ channels, the multijet and Drell-Yan events are
suppressed by requiring $\metrel>45\GeV$. In the $e\mu$ channel, 
Drell-Yan events originate predominantly from $\tau\tau$ production, where the
small leptonic $\tau$ decay branching fractions lead to a much smaller background.
In this channel, the requirement is relaxed to $\metrel>25~\GeV$.
After the isolation and $\metrel$ cuts, the multijet background is found to be
negligible.

Figure~\ref{fig:njets} shows the multiplicity distribution of jets reconstructed using the
anti-$k_{t}$ algorithm~\cite{AntiKt}, with radius parameter $R=0.4$, 
for all events satisfying the pre-selection criteria described above. 
Only jets with $\pt >25\GeV$ and $|\eta| < 4.5$ are considered. This threshold is 
increased to 30~\GeV{} in the region $2.75 < |\eta| < 3.25$, which corresponds
to the boundary between two calorimeter
systems and is more sensitive to reconstruction issues arising from event pile-up.
The background rate and composition depend significantly on jet multiplicity, as
does the signal topology: 
without accompanying jets, the signal originates almost entirely from the ggF
process and the background is dominated by approximately equal fractions of $WW$ and 
Drell-Yan events.
In contrast, when produced in association with two or more jets, the signal
contains a much larger contribution from the VBF process and the background is
dominated by \ttbar{} production.
To maximise the sensitivity, further selection criteria that depend on
the jet multiplicity are applied to the pre-selected sample.
The data are subdivided into 
\ZeroJet, \OneJet{} and \TwoJet{} channels according to the jet counting defined
above, with the \TwoJet{} channel also including higher jet multiplicities.
In addition, slightly different requirements are used for $m_{H} < 200\GeV$,
$200\GeV\leq m_{H} \leq 300\GeV$, and $300\GeV < m_{H} < 600\GeV$; in the following these
are referred to as low $m_{H}$, intermediate $m_{H}$, and high $m_{H}$
selections, respectively. These mass-dependent selections are not mutually exclusive, thus events may contribute to 
more than one mass region. The different requirements for these channels and mass ranges
are described in more detail below.
\begin{figure}[htb!]
  \begin{center}
  \includegraphics[width=0.50\textwidth,bb=16 18 567 407]{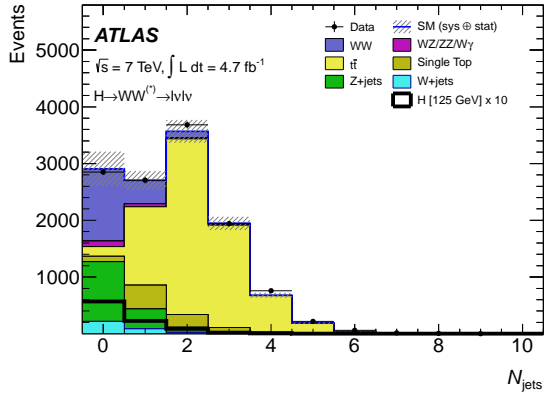}
  \end{center}
  \vspace*{-2mm}
  \caption{Multiplicity of jets within the acceptance described in the text,
    for events satisfying the pre-selection criteria. The lepton flavours are combined.
    The hashed area indicates the total uncertainty on the background prediction.
    The expected signal for a SM Higgs boson with $m_{H}=125\GeV$ is
    superimposed (multiplied by a factor 10 for better visibility).}
  \label{fig:njets}
\end{figure}

%%%%%%%%%%%%%%%%%%%%%%%%%%%%%%%%%%%%%%%%
%          0 jets
%%%%%%%%%%%%%%%%%%%%%%%%%%%%%%%%%%%%%%%%

Due to spin correlations in the $WW^{(\star)}$ system arising from the spin-0 nature of
the Higgs boson, the charged leptons tend to emerge from the interaction point in
the same direction. In the low $m_{H}$ selection this kinematic feature is
exploited for all jet multiplicities by requiring that the azimuthal angular
difference between the leptons, $\Delta\phi_{\ell\ell}$, be less than 1.8
radians, and that the dilepton invariant mass, $\mll$, be less than 50~\GeV{}
for the \ZeroJet{} and \OneJet{} channels. For the \TwoJet{} channel, the $\mll$
upper bound is increased to 80~\GeV{} (the $|\mll - m_{Z}| > 15\GeV$ cut is always applied 
for the same-flavour channels).
For $m_{H} \geq\mhsplit\GeV$, the leptons tend to have higher \pt{} and larger
angular separation.
Therefore, the $\Delta\phi_{\ell\ell}$ cut is omitted and the $\mll$ upper bound
is increased to $150\GeV$.
For $m_{H} > 300\GeV$, the $\mll < 150\GeV$ criterion is also omitted.

In the \ZeroJet{} channel, the magnitude $\ptll$ of the transverse momentum of the 
dilepton system, $\vpTll = {\bf p}_{\rm T}^{\ell 1} + {\bf p}_{\rm T}^{\ell 2}$,
is required to be greater than 30~\GeV{} for the $e\mu$ channel and greater than
45~\GeV{} for the $ee$ and $\mu\mu$ channels.
This improves the rejection of the Drell-Yan background.

%%%%%%%%%%%%%%%%%%%%%%%%%%%%%%%%%%%%%%%%%%%%%%%%%%%%%%%%%%%
%    H+1 Jet
%%%%%%%%%%%%%%%%%%%%%%%%%%%%%%%%%%%%%%%%%%%%%%%%%%%%%%%%%%

In the \OneJet{} channel, backgrounds from top quark decays are suppressed by
rejecting events containing a $b$-tagged jet, as determined using a
$b$-tagging algorithm which uses a combination of impact parameter significance
and secondary vertexing information
and exploits the topology of weak decays of $b$- and $c$-hadrons~\cite{ATLAS-btag-algs}. 
The algorithm is tuned to achieve an 80\% $b$-jet identification 
efficiency in $t\bar{t}$ events while yielding a light-jet tagging rate of approximately
6\%~\cite{ATLAS-CONF-2012-043}.
The total transverse momentum, $\pt^{\rm tot}$,
defined as the magnitude of the vector sum
${\bf p}_{\rm T}^{\rm tot}={\bf p}_{\rm T}^{\ell 1} +
{\bf p}_{\rm T}^{\ell 2}+{\bf p}_{\rm T}^{\rm j}+{\bf p}_{\rm T}^{\rm miss}$,
is required to be smaller than 30~\GeV{} to suppress $t\bar{t}$, single top, and Drell-Yan  
background events with 
jets with $\pt$ below threshold.
The $\tau\tau$ invariant mass, $m_{\tau\tau}$, is computed under the assumption
that the reconstructed leptons are $\tau$ lepton decay products, that the neutrinos
produced in the $\tau$ decays are collinear with the leptons~\cite{CollApp}, and
that they are the only source of \met.
Events in which the computed energies of both putative $\tau$ leptons are
positive (the collinear approximation does not always yield physical solutions)
are rejected if $|m_{\tau\tau}-m_{Z}|<25~\GeV$.

%%%%%%%%%%
%% H+2 Jets
%%%%%%%%%

The \TwoJet{} selection 
follows the \OneJet{} selection described above (with the $\pt^{\rm tot}$
definition modified to include all selected jets).
In addition, the following jet-related cuts are applied:
the two highest-\pt{} jets in the event, the ``tag'' jets, are required to lie 
in opposite pseudorapidity hemispheres ($\eta_{\rm j1}\times\eta_{\rm j2} < 0$), with no additional jet within $|\eta| < 3.2$; 
the tag jets must be separated in pseudorapidity by a distance $|\Delta\eta_{\rm jj}|$
of at least 3.8 units;
finally, the invariant mass of the two tag jets, $m_{\rm jj}$, must be at least 500~\GeV.

A transverse mass variable, \mT~\cite{Barr:2009mx}, is used in this analysis to test for the 
presence of a signal. This variable is defined as: 
\begin{displaymath}
  \label{eq:mT}
  \mT = \sqrt{(E_{\rm T}^{\ell\ell}+\met)^{2} - |\vpTll+{\bf p}_{\rm T}^{\rm miss}|^{2}},
\end{displaymath}
where $E_{\rm T}^{\ell\ell} = \sqrt{|\vpTll|^{2}+m_{\ell\ell}^{2}}$.
The predicted numbers of signal and background events at each stage of the low $m_H$  
selection procedure outlined above are 
presented  in Table~\ref{hww_cutflow}.
Figure~\ref{fig:mT} shows the distributions of the transverse mass after all
 the low $m_H$ selection criteria in the \ZeroJet{} and \OneJet{} analyses, for all lepton
flavours combined. 
No distribution is shown for the \TwoJet{} channel as only
a single event (with $\mT=131\GeV$) is selected in the data.

\begin{figure}[htb!]
  \centering
  \includegraphics[width=0.47\textwidth,bb=12 2 552 400,clip=]{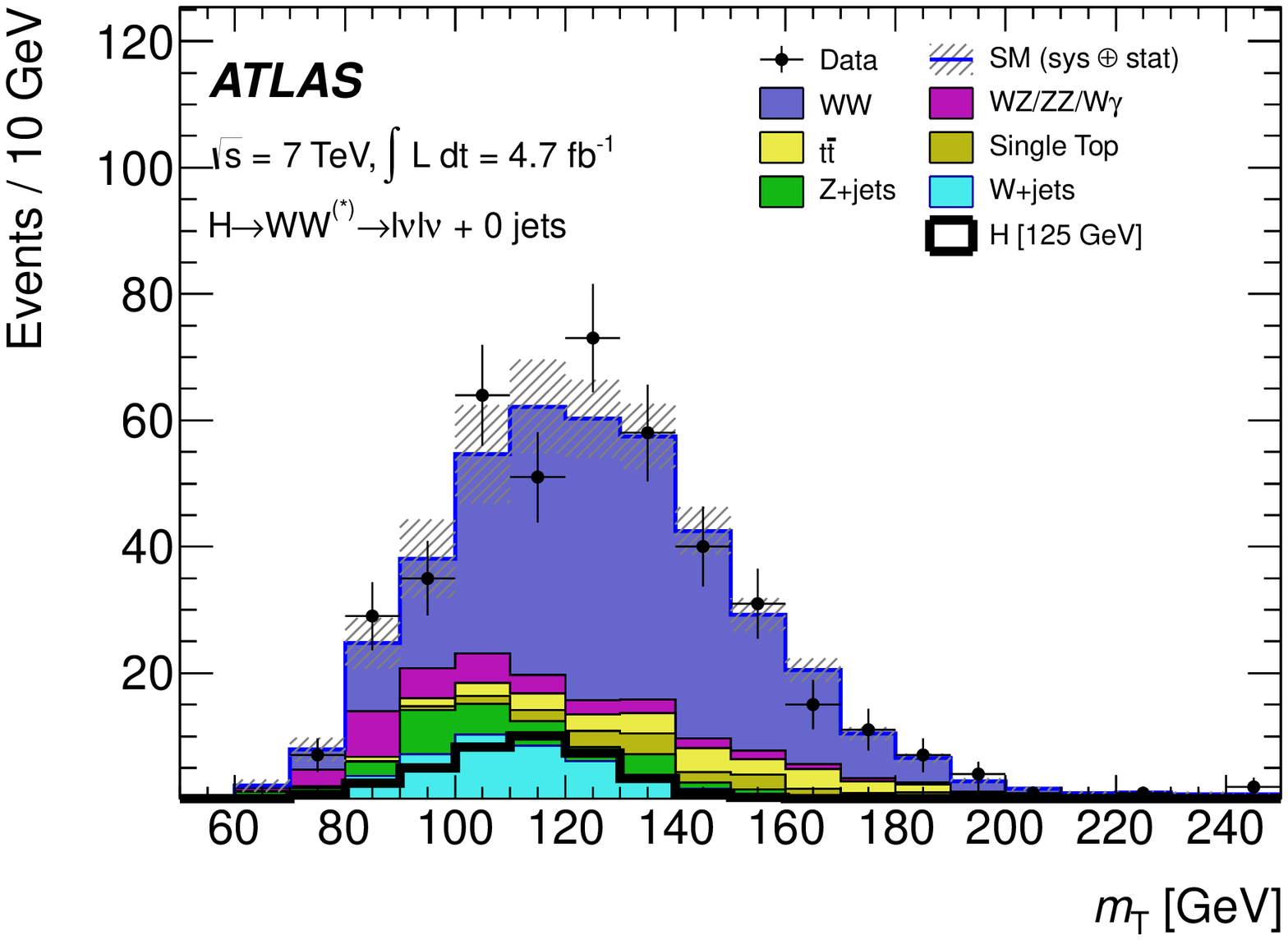}
  \includegraphics[width=0.47\textwidth,bb=12 2 552 400,clip=]{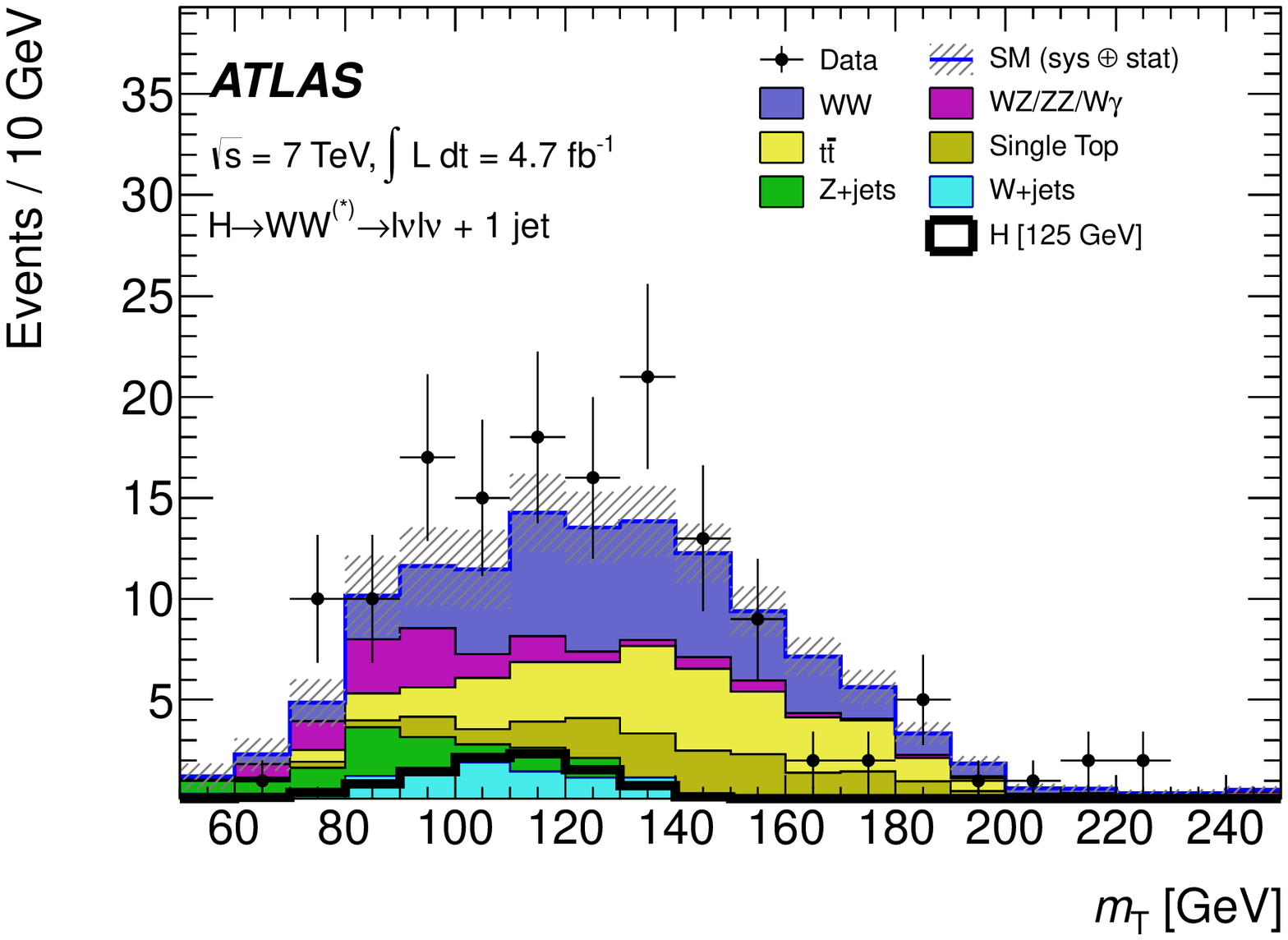}
  \caption{Transverse mass, \mT, distribution in the \ZeroJet{} (top) and
    \OneJet{} (bottom) channels, for events satisfying all criteria
    for the low $m_{H}$ selection. The lepton flavours are combined.
    The expected signal for a SM Higgs boson with $m_{H}=125\GeV$
    is superimposed. % (multiplied by a factor 10 for better visibility).
    The hashed area indicates the total uncertainty on the background prediction.
  }
  \label{fig:mT}
\end{figure}

\section{Background Normalisation and Control Samples}
\label{sec:control}

For the \ZeroJet{} and \OneJet{} analyses, all the main backgrounds from SM
processes producing two isolated high-$\pt$ leptons ($WW$, top, Drell-Yan) are
estimated using partially data-driven techniques based on normalising the MC
predictions to the data in control regions dominated by the relevant background
source.
Only the small background from diboson processes other than $WW$ is estimated
using MC simulation.  For the \TwoJet{} analysis, the $WW$ and Drell-Yan
backgrounds are also estimated using MC simulation.
The backgrounds from fake leptons, which include true leptons from heavy flavour
decays in jets, are fully estimated from data.
The control samples are 
obtained from the data with selections similar to those used in the signal region
but with some criteria reversed or modified to obtain signal-depleted,
background-enriched samples. This helps to reduce the sensitivity of the
background predictions to the systematic uncertainties detailed in
Section~\ref{sec:systematics}. In the following, such control samples are described
for the $WW$, $Z/\gamma^{\star}$+jets, top, and $W$+jets backgrounds.
The quoted uncertainties on the background estimates are those associated with 
the low $m_H$ selection.   

\subsection{$WW$ control sample}
\label{sec:WW-control}

The $WW$ background MC predictions in the \ZeroJet{} and \OneJet{} analyses,
summed over lepton flavours, are normalised using control regions defined with
the same selections as for the signal regions except that the
$\Delta\phi_{\ell\ell}$ requirement is removed.
In addition, the upper selection bound on \mll\ is replaced with a lower bound $\mll >
80~\GeV$ ($\mll > m_{Z} + 15 \GeV$) for the $e\mu$ ($ee$ and $\mu\mu$) final
states.
The numbers of events in the $WW$ control regions in the data  
agree well with the MC predictions, as can be seen in Table~\ref{hww_cutflow}. 
The total uncertainty on the predicted $WW$ background in the signal region is
9\% for the \ZeroJet{} and 22\% for the \OneJet{} analyses.

This control region is used only for the low $m_{H}$ selection in the 
\ZeroJet{} and \OneJet{} analyses. In the intermediate %($200\leq m_{H}\leq 300$)
and high %($m_{H} > 300$)
$m_{H}$ selections, or in the \TwoJet{} analysis, a high-statistics signal-depleted
region cannot be isolated in the data; in these cases, the MC prediction is
used.

\subsection{$Z/\gamma^{\star}$+jets control sample}
\label{sec:DY-control}

In the $ee$ and $\mu\mu$ final states and separately in the \ZeroJet{} and
\OneJet{} analyses, a $Z/\gamma^{\star}$+jets control region is
constructed, after application of all 
selection criteria except that on $\Delta\phi_{\ell\ell}$, by considering a
region with a modified criterion, $20\GeV < \metrel < 45\GeV$.
The number of events in this region, with non-$Z/\gamma^{\star}$+jets
contributions subtracted using the MC prediction, is then scaled by the ratio of events counted in
the $\metrel > 45\GeV$ region to that in the  $20\GeV < \metrel < 45\GeV$ region, for
$|m_{\ell\ell}-m_{Z}| < 15\GeV$. 
Biases in the method are evaluated and corrected for using simulated events.
The acceptance of the $\Delta\phi_{\ell\ell}$ selection criterion is taken from
data.
The resulting uncertainty on the $Z/\gamma^{\star}$+jets background in the
signal region amounts to 38\% and 33\% in the \ZeroJet{} and \OneJet{}
channels, respectively.

In the $e\mu$ channel of the \ZeroJet{} analysis, the background is estimated using the MC simulation and   
cross-checked with data using a control region dominated by $Z\to\tau\tau$ decays, which is 
constructed by requiring $10\GeV < m_{\ell\ell} < 80\GeV$,
$\Delta\phi_{\ell\ell} >$ 2.5, and $\ptll < 30\GeV$.  
A \metrel{} threshold of 25~\GeV{} is used to calculate the data/MC scale factor, 
matching the cut applied to this channel in the signal selection.  The resulting scale factor 
is consistent with unity within the uncertainty of about 10\%.
Owing to the difficulty of
constructing a control region for higher jet multiplicities, a similar 
cross-check cannot be performed for the \OneJet{} and \TwoJet{} analyses. 

\subsection{Top control sample}
\label{sec:top-control}

The estimated number of top quark background events in the \ZeroJet{} signal region
is extrapolated from the number of events
satisfying the pre-selection criteria described in Section~\ref{sec:selection}.
This sample is dominated by top quark backgrounds, as shown in Fig.~\ref{fig:njets}.
The contribution of non-top backgrounds to this sample is subtracted using estimates 
based on MC simulations. 
The scale factor used to propagate the $\ttbar$ contribution in this sample to
the signal region is estimated as the square of the efficiency for one top
quark decay to satisfy the jet veto criterion (estimated using another control
sample, defined by the presence of an additional $b$ jet), with a correction
computed using simulated events to account for single-top background
contributions~\cite{Mellado:2011pc}.
The overall efficiency for the requirements on \ptll, $m_{\ell\ell}$,
and $\Delta\phi_{\ell\ell}$ is taken from simulation.  The total uncertainty on
the top quark background estimate in events with no jets is 22\%.

In the \OneJet{} and \TwoJet{} analyses, the top quark background MC prediction is
normalised to the data using a control sample defined by reversing the $b$-jet
veto and removing the requirements on $\Delta\phi_{\ell\ell}$ and
$m_{\ell\ell}$.
The resulting samples are dominated by top quark backgrounds (both \ttbar{} and
single-top production), with little contribution from other sources.
Good agreement between data and MC for the numbers of events in the \OneJet{} and \TwoJet{}  
control regions is observed (see Table Table~\ref{hww_cutflow}).
%The numbers of events in the \OneJet{} and \TwoJet{}
%control regions are given in Table~\ref{hww_cutflow}. 
The total uncertainties on
the estimated top quark background in the \OneJet{} and \TwoJet{} signal regions
amount to 23\% and 40\%, respectively.

\subsection{$W$+jets control sample}
\label{sec:wjets-control}

The $W$+jets background contribution is estimated using a data 
sample of events where one of the two leptons satisfies the identification and
isolation criteria described in Section~\ref{sec:selection}, and the other
lepton (denoted ``anti-identified'') fails these criteria while satisfying a
loosened selection.
All other selection criteria follow those applied in the signal region.
The dominant contribution to this background comes from $W$+jets production with jets faking electrons. 
The contamination in the signal region is then
obtained by scaling the number of events in the data control
sample by a normalisation ``fake factor''.  
The fake factor is estimated as a function of the anti-identified lepton $\pt$ using
an inclusive dijet data sample, after subtracting the residual contributions from real leptons arising 
from leptonic $W$ and $Z$ decays.  
The $W$ candidates are identified by 
requiring the transverse mass
$\mT^{W} = \sqrt{ 2 \pt^\ell \met \cdot (1 -\cos\Delta\phi)}$ to
satisfy $\mT^{W} > 30\GeV$. In this expression, $\pt^\ell$ is the lepton
transverse momentum and $\Delta\phi$ is the difference in azimuth between the
lepton and missing transverse momentum directions.
The $Z$ candidates are identified by requiring two opposite-sign leptons of the
same flavour and $|m_{\ell\ell} -m_{Z}| < 15\GeV$.
The small remaining lepton contamination, which includes $W\gamma$ and $W\gamma^{\star}$ events, is subtracted using MC simulation.   
The fake factor uncertainty is the main uncertainty on the $W$+jets background
contribution.
This uncertainty is dominated by differences in jet properties between dijet and
$W$+jets samples evaluated with simulated events, with smaller contributions
originating from trigger effects and the subtraction of the contamination from
real leptons from leptonic $W$ and $Z$ decays.
The total uncertainty on this background is estimated to be
approximately 60\%.

\section{Systematic Uncertainties}
\label{sec:systematics}

Theoretical uncertainties on the signal production cross sections
are determined following 
Refs.~\cite{LHCHiggsCrossSectionWorkingGroup:2011ti,LHCHiggsCrossSectionWorkingGroup:2012vm}.
QCD renormalisation %($\mu_{\mathrm{R}}$)
and factorisation scales %($\mu_{\mathrm{F}}$)
are varied up and down independently by a factor of two.
Independent uncertainties on the
ggF signal production are assumed for the inclusive cross section and the cross
section for production with at least one or two jets. The resulting
uncertainties on the cross sections in exclusive jet multiplicity analyses
are taken into account, as well as anti-correlations caused by 
transitions between jet multiplicities. The relative \ZeroJet{} (\OneJet)
cross section uncertainties depend on $m_{H}$, rising from $\pm$21\% ($\pm$31\%)
at $m_{H}=125\GeV$ and $m_{H}=240\GeV$ to $\pm$42\% ($\pm$31\%) at
$m_{H}=600\GeV$~\cite{Stewart:2011cf,ATL-PHYS-PUB-2011-011,LHCHiggsCrossSectionWorkingGroup:2012vm}.
The \TwoJet{} analysis is mainly sensitive to the VBF process. The impact of the
scale variations on the combined VBF signal cross section and jet veto
acceptance is 4\%~\cite{LHCHiggsCrossSectionWorkingGroup:2012vm}.
In this analysis, around 25\% of the signal events are produced via
ggF, where the relative uncertainty is around 25\%.
For the high mass range, an additional uncertainty due to the Higgs lineshape description in the POWHEG MC generator is added 
in quadrature for both the ggF and the VBF channel and amounts to $150\% \times
(m_{H}/1\TeV)^{3}$~\cite{Seymour:1995qg,Passarino:2010qk,Anastasiou:2011pi,LHCHiggsCrossSectionWorkingGroup:2012vm}.
The uncertainties associated with the underlying event and parton showering are taken into
account in the acceptance uncertainty, but they are negligible compared to
the scale uncertainties on the cross sections in exclusive jet bins.

PDF uncertainties are estimated, following Refs.~\cite{Botje:2011sn,Lai:2010vv,Martin:2009iq,Ball:2011mu},
by the envelopes of error sets as well as different PDF sets, applied separately to quark-quark,
quark-gluon, and gluon-gluon initiated processes. The relative PDF uncertainty
on the dominant ggF signal process is about 8\%; the VBF uncertainty varies from
$\pm 2$\% at $m_{H}=125\GeV$ to $\pm 4$\% at $m_{H}=600\GeV$.
Uncertainties on the modelling of signal and background processes are estimated
by using alternative generators, such as MC@NLO for the ggF process, 
ALPGEN for $WW$ production, POWHEG for $t\bar{t}$ production, and PYTHIA for the 
 $Z/\gamma^{\star}$+2 jets process. 
The uncertainties associated with the underlying event and parton showering are taken into 
account in the acceptance uncertainty, but they are negligible compared to
the scale uncertainties on the cross sections in exclusive jet bins.

The main experimental uncertainties are related to the jet energy scale which is
determined from a combination of test beam, simulation, and \emph{in situ}
measurements.
The uncertainty on the jet energy scale varies from 14\% to 2\% as a function of
jet $\pt$ and $\eta$ for jets with 
$\pt > 25\GeV$ and $|\eta| < 4.5$~\cite{JES}; for central jets it is at most 4\%.
An additional contribution from event pile-up is estimated to vary between 5\% and
0.5\%, depending on jet $\pt$ and $\eta$, for jets with $\pt > 25\GeV$.
The uncertainty on the jet energy resolution is estimated from \emph{in situ} measurements.
The resolution varies from 25\% to 5\%, and its uncertainty from 5\% to
2\%, as a function of jet $\pt$ and $\eta$.
The reconstruction, identification, and trigger efficiencies for electrons and muons, as well as their momentum scales
and resolutions, are estimated 
using $Z\to\ell\ell$, $J/\psi\to\ell\ell$, and $W\to\ell\nu$ decays. % ($\ell = e$, $\mu$).
With the exception of the uncertainty on the electron efficiency, which
varies between 2\% and 5\% as a function of \pt{} and $\eta$,
the resulting uncertainties are all smaller than 1\%.
Jet energy scale and lepton momentum scale uncertainties are propagated to the
\met{} computation. Additional contributions arise from jets with $\pt < 20\GeV$
as well as from low-energy calorimeter deposits not associated with reconstructed
physics objects~\cite{MET}; %, jointly referred to as low-energy deposits;
their effect on the total background event yield ranges from 1\% to  8\%.
Finally, uncertainties on the modelling of event pile-up contributions are estimated by
varying their effect on low-energy calorimeter deposits;
the impact on the background yield varies between 1\% and 5\%. 
The efficiency of the $b$-tagging algorithm is calibrated
using samples containing muons reconstructed in the vicinity of
jets~\cite{ATLAS-CONF-2012-043}. The resulting uncertainty on the $b$-jet tagging
efficiency varies between 5\% and 14\% as a function of jet \pt.
The uncertainty on the integrated luminosity is
3.9\%~\cite{luminosity-paper,ATLAS-CONF-2011-116}. 

In this analysis, a fit to the \mT{} distribution is performed in order to obtain 
the signal yield for each mass hypothesis. 
The \mT{} shapes for the individual backgrounds and signal
do not exhibit a statistically significant
dependence on the majority of the theoretical and experimental uncertainties.
The remaining uncertainties that do produce statistically significant variations
of the \mT{} shape have no appreciable effect on the final results.
Hence, the uncertainty on the shape of the total background is dominated by the
uncertainties on the normalisations of the individual backgrounds.

Systematic uncertainties are evaluated for the control regions described in
Section~\ref{sec:control} in the same fashion as for the signal region.
For the backgrounds normalised using these control regions, only the relative
normalisation between the backgrounds in the signal and control regions is
affected.

\section{Results}
\label{sec:results}

The expected numbers of signal ($m_{H}=125\GeV$) and
background events at several stages of the low $m_{H}$ selection 
are presented in Table~\ref{hww_cutflow}.
The rightmost column shows the observed numbers of events in the data. 
The uncertainties shown include only the statistical uncertainties on the
predictions from simulation and on the normalisation of the dominant backgrounds.
After all selection criteria, the dominant background 
in the \ZeroJet{} channel comes from continuum $WW$ production, with 
smaller contributions from top ($t\bar{t}$ and single top) and $W$+jets events.
In the \OneJet{} and \TwoJet{} channels, the $WW$ and top backgrounds are
comparable.

\begin{table*}[!htbp]
\begin{center}
\caption{
The expected numbers of signal and background events after the requirements
of the low $m_{H}$ selection
listed in the first column, %in 4.7~fb$^{-1}$ of integrated luminosity, 
as well as the observed numbers of events.
The signal is for $m_{H} = 125\GeV$.
The $W$+jets background is estimated entirely from data, whereas
MC predictions normalised to data in control regions are used
for the $WW$, $Z/\gamma^{\star}$+jets, $t\bar{t}$,
and $tW/tb/tqb$ processes.  Contributions
from other background sources are taken from MC predictions.
Only statistical uncertainties associated with the number of events in 
the MC samples and in the data control regions are shown.
The expected numbers of signal and background events, and the observed
numbers of events, are shown also in the control regions; here, with the exception
of $W$+jets, no normalisation scale factors are applied to the expected
background contributions.
The bottom part of the table 
lists the number of expected and observed events for each lepton channel after the $\Delta\phi_{\ell\ell}$ cut.
}
\label{hww_cutflow}
\vspace*{0.2cm}
\resizebox{\textwidth}{!}{
\begin{tabular}{l|r@{$\,\pm \,$}l r@{$\,\pm \,$}l r@{$\,\pm \,$}l r@{$\,\pm \,$}l r@{$\,\pm \,$}l r@{$\,\pm \,$}l r@{$\,\pm \,$}l r@{$\,\pm \,$}l |c}
\hline
\hline
    \parbox{0pt}{~\\\vspace{0.45cm}}      \ZeroJet & \multicolumn{2}{c}{Signal} & \multicolumn{2}{c}{$WW$} &  \multicolumn{2}{c}{$WZ/ZZ/W\gamma$} & \multicolumn{2}{c}{$t\bar{t}$} & \multicolumn{2}{c}{$tW/tb/tqb$} &\multicolumn{2}{c}{$Z/\gamma^{\star}+\mathrm{jets}$} &  \multicolumn{2}{c}{$W+\mathrm{jets}$} &   \multicolumn{2}{c}{Total Bkg.} & Obs. \\
\hline
%Jet Veto                                 & 56.7 & 0.2     & 1273 & 79   &  97 &  4   & 174 & 12   & 95  & 7   & 1039  & 28   & 217  & 4   & 2893 & 115   & 2849\\
Jet Veto                                 & 56.7 & 0.2     & 1273 & 79   &  97 &  4   & 174 & 12   & 95  & 7   & 1039  & 28   & 217  & 4   & 2890 & 120   & 2849\\
$m_{\ell\ell} < 50\GeV$              & 45.2 & 0.2     & 312 & 20      & 41 & 3     & 29  & 2    & 19  & 2   & 168     & 10    & 70  & 2       & 639 & 28    & 645\\
$p_{\rm T}^{\ell\ell}$ cut             & 40.1 & 0.2     & 282 & 18   & 35 & 3    & 28  & 2     & 18   &  2   & 28 & 6 & 49 & 2 & 439 & 26    & 443\\
$\Delta\phi_{\ell\ell} < 1.8$            & 39.0 & 0.2     & 276  & 17   & 33 & 2    & 27   & 2     & 18   &  2   & 28 & 6 & 44 & 1 & 425 & 26    & 429\\
\hline
  \parbox{0pt}{~\\\vspace{0.45cm}}      \OneJet  & \multicolumn{2}{c}{Signal} & \multicolumn{2}{c}{$WW$} &  \multicolumn{2}{c}{$WZ/ZZ/W\gamma$} & \multicolumn{2}{c}{$t\bar{t}$} & \multicolumn{2}{c}{$tW/tb/tqb$} &\multicolumn{2}{c}{$Z/\gamma^{\star}+\mathrm{jets}$} &  \multicolumn{2}{c}{$W+\mathrm{jets}$} &   \multicolumn{2}{c}{Total Bkg.} & Obs. \\
\hline
1 jet                                    & 22.7 & 0.1 & 343 & 54 & 56 & 3 & 1438 & 60 & 436 & 19 & 357 & 17 & 85 & 3 & 2720 & 140 & 2706 \\
$b$-jet veto                             & 20.9 & 0.1 & 319 & 50 & 52 & 3 & 412 & 18 & 139 & 7 & 332 & 16  & 76 & 3 & 1330 & 84 & 1369  \\
$|{\bf p}_{\rm T}^{\rm tot}| < 30\GeV$    & 14.0 & 0.1 & 226 & 35 & 34 & 2 & 181 & 8 & 80 & 4 & 108 & 8  & 37 & 2 & 666 & 51 & 684  \\
$Z\rightarrow\tau\tau$ veto    & 14.0 & 0.1 & 220 & 34 & 34 & 2 & 173 & 8 & 77 & 4 & 85 & 7  & 37 & 2 & 627 & 50 & 644  \\
$m_{\ell\ell} < 50\GeV$             & 10.9 & 0.1 & 49 & 8 & 14 & 2 & 33 & 2 & 18 & 1 & 24 & 3 & 12 & 1 & 148 & 12 & 170  \\
$\Delta\phi_{\ell\ell} < 1.8$       & 10.1 & 0.1 & 44 & 7 & 13 & 2 & 31 & 2 & 17 & 1 & 10 & 2 & 10 & 1 & 126 & 10 & 145  \\     
\hline
 \parbox{0pt}{~\\\vspace{0.45cm}}      \TwoJet  & \multicolumn{2}{c}{Signal} & \multicolumn{2}{c}{$WW$} & \multicolumn{2}{c}{$WZ/ZZ/W\gamma$} & \multicolumn{2}{c}{$t\bar{t}$} & \multicolumn{2}{c}{$tW/tb/tqb$} &\multicolumn{2}{c}{$Z/\gamma^{\star}+\mathrm{jets}$} &  \multicolumn{2}{c}{$W+\mathrm{jets}$} &   \multicolumn{2}{c}{Total Bkg.} & Obs. \\
\hline
$\ge 2$ jets & 11.4 & 0.1 & 142 & 2 & 26 & 2 & 5939 & 17 & 339 & 5 & 120 & 7 & 40 & 4 & 6605 & 20 & 6676  \\
Central jet veto & 9.0 & 0.1 & 113 & 2  & 20 & 1 & 3279 & 13 & 238 & 4 & 89  & 6 & 25 & 3 & 3765 & 15 & 3811  \\
$b$-jet veto & 7.6 & 0.1 & 98 & 1  & 18 & 1 & 353 & 4 & 51 & 2 & 77 & 5  & 19 & 2 & 615 & 8 & 667   \\
Opp. hemispheres & 4.2 & 0.1 & 46 & 1  & 7 & 1 & 149 & 3 & 21 & 1 & 32 & 3  & 9 & 1 & 264 & 5 & 269 \\
$|\Delta\eta_{\mathrm{jj}}|>3.8$ & 1.8 & 0.1 & 8.4 & 0.4 & 0.9 & 0.2 & 23.2 & 1.0 & 2.2 & 0.4 & 5.8 & 1.7  & 1.7 & 0.4 & 42.2 & 2.1 & 40  \\
 $m_{\mathrm{jj}}>500\GeV$ & 1.3 & 0.1 & 3.9 & 0.3 & 0.4 & 0.1 & 10.4 & 0.6 & 1.0 & 0.3 & 0.7 & 0.4  & 0.9 & 0.3 & 17.3 & 0.9 & 13  \\
$m_{\ell\ell} < 80\GeV$              & 0.9 & 0.1     & 1.1 & 0.2      & 0.1 & 0.1     & 1.4  & 0.2    & 0.4  & 0.1   & 0.2     & 0.2    & 0.2  & 0.2       & 3.2 & 0.4    & 2\\
 $\Delta\phi_{\ell\ell}<1.8$ & 0.8 & 0.1 & 0.8 & 0.1 & 0.1 & 0.1 & 0.9 & 0.2 & 0.1 & 0.1 &  \multicolumn{2}{c}{negl.}  & \multicolumn{2}{c}{negl.} & 1.8 & 0.3 & 1  \\
\hline
\hline
\parbox{0pt}{~\\\vspace{0.45cm}}  Control Regions      & \multicolumn{2}{c}{Signal} & \multicolumn{2}{c}{$WW$} &  \multicolumn{2}{c}{$WZ/ZZ/W\gamma$} & \multicolumn{2}{c}{$t\bar{t}$} & \multicolumn{2}{c}{$tW/tb/tqb$} &\multicolumn{2}{c}{$Z/\gamma^{\star}+\mathrm{jets}$} &  \multicolumn{2}{c}{$W+\mathrm{jets}$} &   \multicolumn{2}{c}{Total Bkg.} & Obs. \\
\hline
$WW$ 0-jet & 0.3 & 0.1 & 471 & 3 & 26 & 1 & 87 & 2 & 42 & 2 & 7 & 2  & 49 & 2 & 682 & 5 & 697  \\
$WW$ 1-jet   & 0.1 & 0.1 & 128 & 2 & 12 & 1 & 89 & 2 & 34 & 2 & 9 & 2 & 11 & 1 & 282 & 4 & 270  \\
Top 1-jet & 1.2 & 0.1 & 20 & 1 & 1.9 & 0.5 & 434 & 4 & 169 & 4 & 7 & 2  & 4 & 1 & 635 & 6 & 676  \\
Top 2-jet & 0.1 & 0.1 & 0.4 & 0.1 & \multicolumn{2}{c}{negl.} & 10.0 & 0.7 & 1.0 & 0.3 & \multicolumn{2}{c}{negl.} & \multicolumn{2}{c}{negl.} & 11.4 & 0.7 & 10\\
\hline \hline 
\end{tabular}
}
\scalebox{0.8}{
\begin{tabular}{l|r@{$\,\pm \,$}l r@{$\,\pm \,$}l r@{$\,\pm \,$}l r@{$\,\pm \,$}l r@{$\,\pm \,$}l r@{$\,\pm \,$}l r@{$\,\pm \,$}l r@{$\,\pm \,$}l |c}
\hline
\parbox{0pt}{~\\\vspace{0.45cm}}  Lepton Channels           & \multicolumn{2}{c}{0-jet $ee$} & \multicolumn{2}{c}{0-jet $\mu\mu$} & \multicolumn{2}{c}{0-jet $e\mu$} & \multicolumn{2}{c}{1-jet $ee$} & \multicolumn{2}{c}{1-jet $\mu\mu$} & \multicolumn{2}{c}{1-jet $e\mu$}              \\
      \cline{1-13}
      Total bkg. & 60 &   5 & 116 &  10 &  249 &  12 & 19  &   2 &  34 &   4 &  72  & 6  \\
      Signal    & 4.0 & 0.1 & 9.4 & 0.1 & 25.7 & 0.2 & 1.2 & 0.1 & 2.5 & 0.1 & 6.4  & 0.1   \\
      \cline{1-13}
      Observed &               \multicolumn{2}{c}{52}   &
      \multicolumn{2}{c}{138}  & \multicolumn{2}{c}{239} &
      \multicolumn{2}{c}{19} & \multicolumn{2}{c}{36} & \multicolumn{2}{c}{90}  \\     
  \hline \hline
\end{tabular}
}
\end{center}
\end{table*}

Table~\ref{hww_comparison} shows the numbers of events expected from signal and background
and observed in data, after application of all selection criteria.
To reflect better the sensitivity of the analysis, an additional mass-dependent
cut on $\mT$ has been applied: $0.75\,m_{H}<m_{\rm T}<m_{H}$ for $m_{H}=125\GeV$ and
$0.6\,m_{H}<m_{\rm T}<m_{H}$ for $m_{H}=240\GeV$.
The uncertainties shown in Table~\ref{hww_comparison} include those of Table~\ref{hww_cutflow}
as well as the systematic uncertainties discussed in
Section~\ref{sec:systematics}, constrained by the use of the control
regions discussed in Section~\ref{sec:control}.  The uncertainties are those that 
enter into the fitting procedure described below.
Table~\ref{hww_systematics} shows the effect of the main sources of systematic
uncertainty on the signal ($m_{H}=125\GeV$) and background predictions 
for the three jet multiplicity analyses. Similarly to Table~\ref{hww_comparison}, the additional \mT{} cut is
applied and the constraints from control regions are included.

\begin{table*}[!htbp]
\begin{center}
\caption{
The expected numbers of signal ($m_{H}=125\GeV$ and 240~\GeV) and background
events after the full low $m_{H}$ and intermediate $m_{H}$ selections, 
including a cut on the transverse mass of 
$0.75\,m_{H}<m_{\rm T}<m_{H}$ for $m_{H}=125\GeV$ and
$0.6\,m_{H}<m_{\rm T}<m_{H}$ for $m_{H}=240\GeV$.
The observed numbers of events are also displayed.
The uncertainties shown are the combination of the statistical and all
systematic uncertainties, taking into account the constraints from control
samples. These results and uncertainties differ from those given in Table~\ref{hww_cutflow} 
due to the application of the additional $\mT$ cut.
All numbers are summed over lepton flavours. 
}
\label{hww_comparison}
\vspace*{0.2cm}
\resizebox{\textwidth}{!}{
\begin{tabular}{cc|r@{$\,\pm \,$}l r@{$\,\pm \,$}l r@{$\,\pm \,$}l r@{$\,\pm \,$}l r@{$\,\pm \,$}l r@{$\,\pm \,$}l r@{$\,\pm \,$}l r@{$\,\pm \,$}l |c}
\hline
\hline
    \parbox{0pt}{~\\\vspace{0.45cm}}   &     & \multicolumn{2}{c}{Signal} & \multicolumn{2}{c}{$WW$} &  \multicolumn{2}{c}{$WZ/ZZ/W\gamma$} & \multicolumn{2}{c}{$t\bar{t}$} & \multicolumn{2}{c}{$tW/tb/tqb$} &\multicolumn{2}{c}{$Z/\gamma^{\star}+\mathrm{jets}$} &  \multicolumn{2}{c}{$W+\mathrm{jets}$} &   \multicolumn{2}{c}{Total Bkg.} & Obs. \\
\hline
\multirow{2}{0.1cm}{\begin{sideways}$0$-jet\end{sideways}} &
     $m_H$ = 125 GeV & 26 &  7 & 108 & 12 & 12 & 2 &  7 & 2  & 5  & 1 & 14 & 6 & 27 & 16 & 172 & 21 & 174 \\
   & $m_H$ = 240 GeV & 61 & 16 & 450 & 48 & 24 & 3 & 73 & 15 & 42 & 9 & 6  & 3 & 36 & 24 & 632 & 64 & 627 \\
\hline
 \multirow{2}{0.1cm}{\begin{sideways}$1$-jet\end{sideways}} &
   $m_{H}$ = 125 GeV &  6 &  2 & 16  &  3 &  5 & 2 &  8 &  2   &   4 &   2 & 5 & 2 & 5 & 3 &  42 &  6 &  56\\
 & $m_{H}$ = 240 GeV & 24 &  8 & 95  & 20 &  9 & 1 & 84 & 23   &  39 &  16 & 5 & 2 & 8 & 7 & 241 & 48 & 232\\
\hline
 \multirow{2}{0.1cm}{\begin{sideways}$2$-jet\end{sideways}} &
   $m_{H}$ = 125 GeV & 0.5 & 0.1 & 0.2 & 0.2 & \multicolumn{2}{c}{negl.} & 0.2 & 0.1 & \multicolumn{2}{c}{negl.} & \multicolumn{2}{c}{$0.0^{+0.1}_{-0.0}$\phantom{0000}}  & \multicolumn{2}{c}{negl.} & 0.4 & 0.3 & 0  \\
 & $m_{H}$ = 240 GeV & 2.6 & 0.4 & 1.2 & 0.8 & 0.1 & 0.1 & 2.2 & 1.0 & 0.3 & 0.2 & \multicolumn{2}{c}{negl.}  & 0.1 & 0.1 & 3.9 & 1.5 & 2  \\
\hline
\hline
\end{tabular}
}
\end{center}
\end{table*}

The statistical analysis of the data employs a binned likelihood function
$\mathcal{L}(\mu,\boldsymbol{\theta})$ constructed as the product of Poisson
probability terms in each lepton flavour channel. The mass-dependent cuts on
\mT{} described above are not used.
Instead, the \ZeroJet{} (\OneJet) signal regions are subdivided into five (three)
\mT{} bins. For the \TwoJet{} signal region (where the small number of events
remaining after the selection does not allow the use of shape information), and
for the $WW$ and top control regions, only the results integrated over \mT{} are used.
Because of event pile-up conditions changing throughout data-taking and
leading to a progressively worsening \met{} resolution, separate likelihood
terms are constructed (both for the signal and the control
regions) for the first 2.3~\ifb{} and the remaining 2.4~\ifb{} dataset.
A ``signal strength'' parameter, $\mu$, multiplies the expected Standard Model
Higgs boson production signal in each bin.
Signal and background predictions depend on systematic uncertainties that are
parameterised by nuisance parameters $\boldsymbol{\theta}$, which in turn are
constrained using Gaussian functions.
The expected signal and background event counts in each bin are functions of
$\boldsymbol{\theta}$. The parameterisation is chosen such that the rates in each
channel are log-normally distributed for a normally distributed $\theta$.
The test statistic $q_\mu$ is then constructed using the profile likelihood:
$q_\mu = -2\ln\left(\mathcal{L}(\mu,\hat{\boldsymbol{\theta}}_\mu)/
                   \mathcal{L}(\hat{\mu},\hat{\boldsymbol{\theta}})\right)$,
where $\hat{\mu}$ and $\hat{\boldsymbol{\theta}}$ are the parameters that
maximise the likelihood (with the constraint $0 \leq \hat{\mu} \leq \mu$), and
$\hat{\boldsymbol{\theta}}_\mu$ are the nuisance parameter values that maximise
the likelihood for a given $\mu$.
This test statistic is used to compute exclusion limits following the modified
frequentist method known as $CL_s$~\cite{CLs_2002,asymptotics}.

\begin{table}[htbp]
\centering
\caption{
Main relative systematic uncertainties on the predicted numbers of signal ($m_{H}=125\GeV$) and
background events for each of the three jet multiplicity analyses.
The same $\mT$ criteria as in Table~\ref{hww_comparison} are imposed in addition
to the low $m_{H}$ signal selection criteria. All numbers are summed over lepton
flavours. The effect of the quoted inclusive signal cross section renormalisation and factorisation scale uncertainties
on exclusive jet multiplicities is explained in Section~\ref{sec:systematics}.
}
\label{hww_systematics}
\vspace*{0.2cm}
\resizebox{0.48\textwidth}{!}{
  \begin{tabular}{l|rr}
    \hline
    \hline
    \parbox{0pt}{~\\\vspace{0.30cm}} Source (0-jet) & Signal (\%) & Bkg. (\%)\\
    \hline
    Inclusive ggF signal ren./fact. scale   & 19 & 0 \\
    1-jet incl. ggF signal ren./fact. scale & 10 & 0 \\
    $W$+jets fake factor                    & 0 & 10 \\
    Parton distribution functions           & 8 &  2 \\
    $WW$ normalisation                      & 0 &  6 \\
    Jet energy scale                        & 6 &  0 \\
    \hline
    \parbox{0pt}{~\\\vspace{0.30cm}} Source (1-jet) & Signal (\%) & Bkg. (\%)\\
    \hline
    1-jet incl. ggF signal ren./fact. scale & 27 & 0 \\
    2-jet incl. ggF signal ren./fact. scale & 15 & 0 \\
    Missing transverse momentum   &  8 & 3 \\
    $W$+jets fake factor          &  0 & 7 \\
    $b$-tagging efficiency        &  0 & 7 \\
    Parton distribution functions &  7 & 1 \\
    \hline
    \parbox{0pt}{~\\\vspace{0.30cm}} Source (2-jet) & Signal (\%) & Bkg. (\%)\\
    \hline
    Jet energy scale                       & 13 & 36 \\
    $Z/\gamma^{\star}$+2 jets MC modelling &  0 & 24 \\ 
    Diboson ren./fact. scale               &  0 & 22 \\
    \hline
    \hline
  \end{tabular}
}
\end{table}

Figure~\ref{fig:LimitCls} shows, the observed and expected cross section upper
limits at 95\% CL, as a function of $m_{H}$ and normalised to the SM cross 
section, for the combined \ZeroJet, \OneJet{} and \TwoJet{} analyses.
The limits exclude a Standard Model Higgs boson with a mass in the range from
$\limobslow\GeV$ to $\limobshigh\GeV$ at 95\% CL, while the expected exclusion
range in the absence of a signal is $\limexplow\GeV \leq \mH \leq \limexphigh\GeV$.
No significant excess of events over the expected background is observed over
the entire mass range (the lowest $p$-value observed is 0.15).

\begin{figure}[hbt!]
  \centering
  \includegraphics[width=0.47\textwidth,bb=22 12 522 410,clip=]{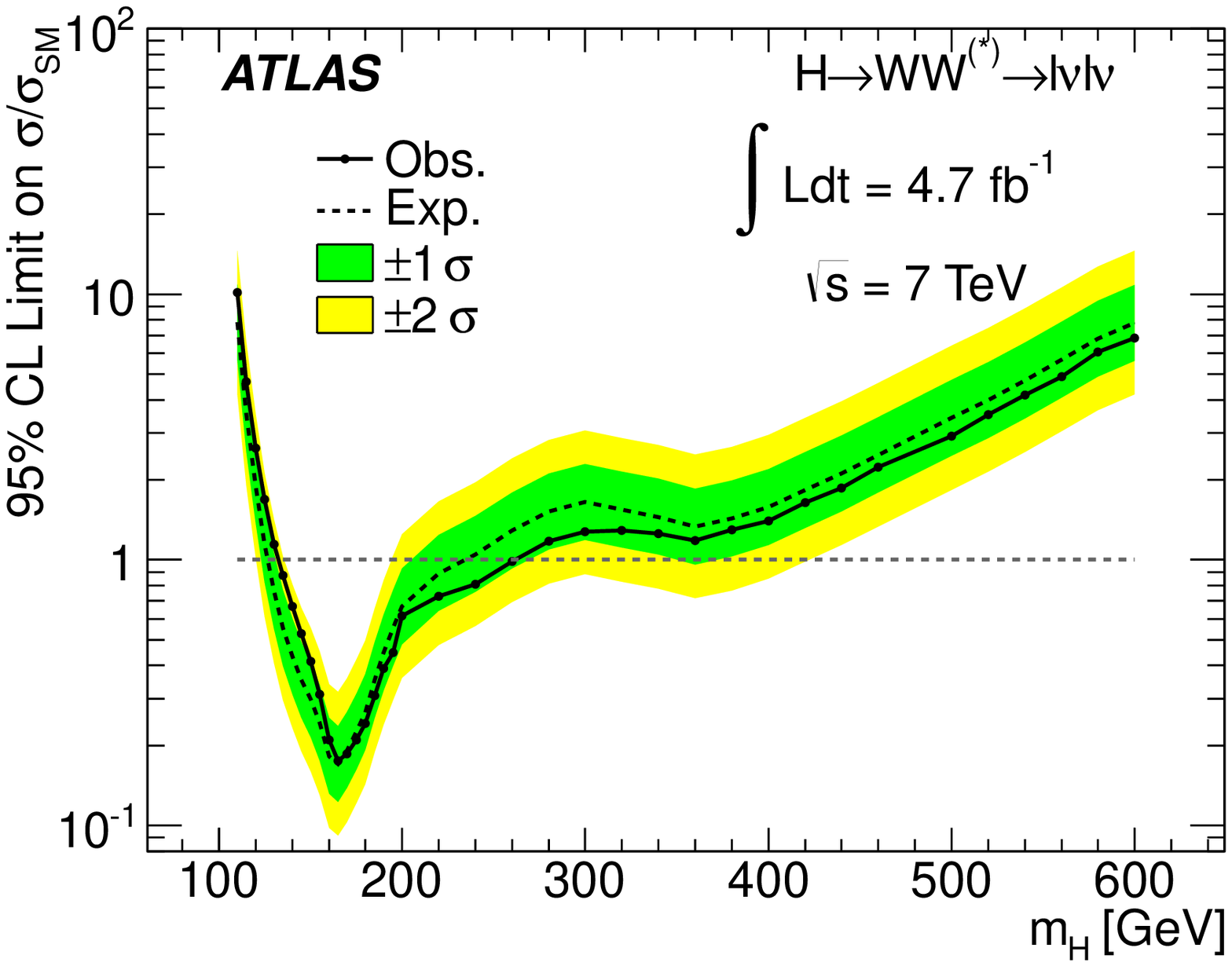}
  \includegraphics[width=0.47\textwidth,bb=22 12 522 410,clip=]{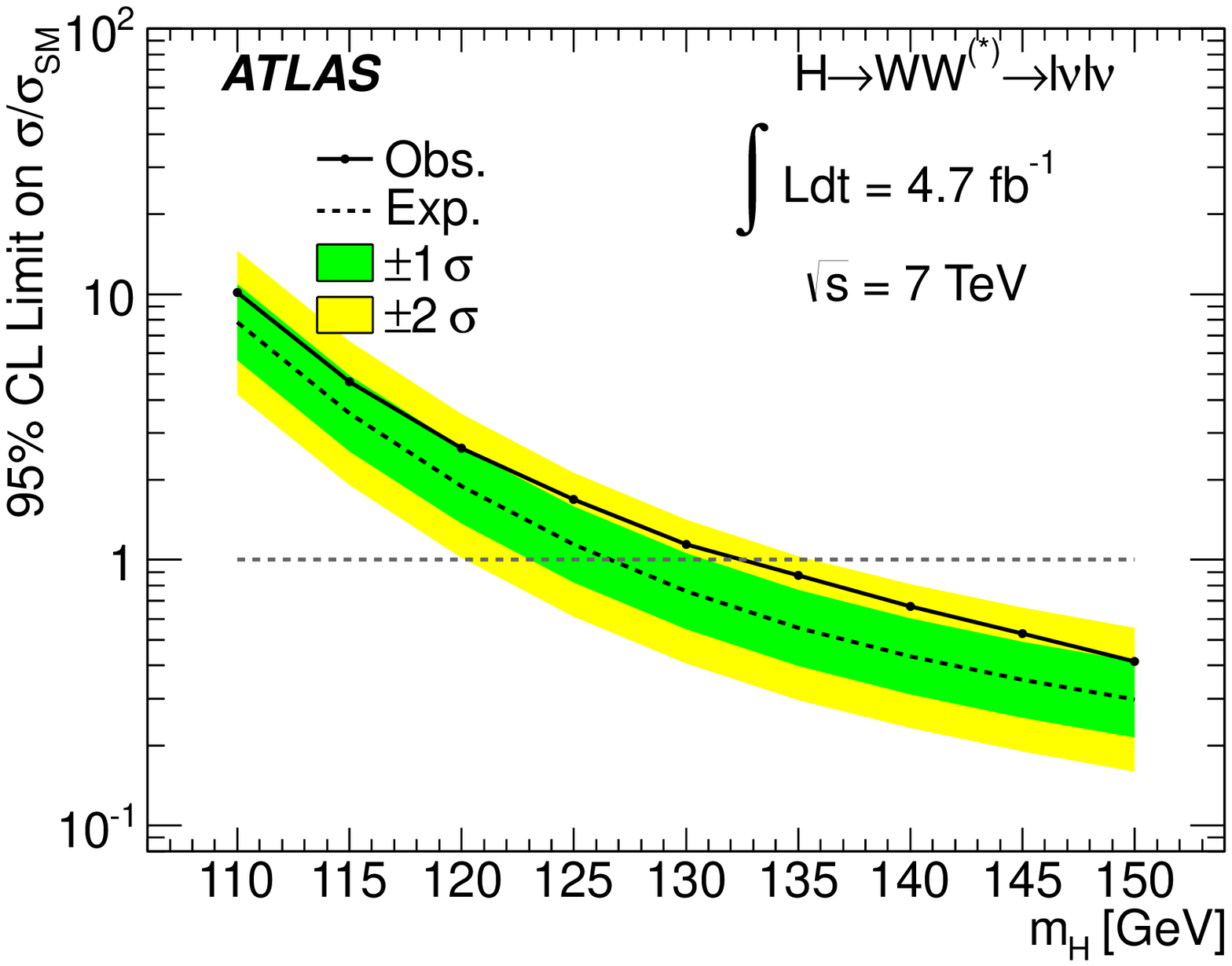}
  \vspace*{-0.5cm}
  \caption{Observed (solid) and expected (dashed) 95\% CL upper limits on the
    Higgs boson production cross section, normalised to the SM cross section, as a function of $m_{H}$,
    over the full mass range considered in this analysis (top) and restricted to
    the range $m_{H} < 150\GeV$ (bottom).
    The inner (green) and outer (yellow) regions indicate the $\pm 1\sigma$ and $\pm 2\sigma$
    uncertainty bands on the expected limit, respectively.
    The results for nearby masses are highly correlated due to the
    limited mass resolution (5--8~\GeV, as inferred from a study of the effect
    of a hypothetical $m_{H}=125\GeV$ signal on the behaviour of
    $q_{\mu}(\mu=1)$ as a function of $m_{H}$) in this final state.}
  \label{fig:LimitCls}
\end{figure}

\section{Conclusion}

A search for the SM Higgs boson has been performed in the \hwwlnln\ channel
using the full data sample (4.7 fb$^{-1}$) of $pp$ collision data from the Large
Hadron Collider at $\sqrt{s} = 7\TeV$ recorded in 2011 with the ATLAS detector.
No significant excess of events over the expected background is observed.
A SM Higgs boson with mass in the range from \limobslow~\GeV{} to \limobshigh~\GeV{} is excluded at 95\% CL, 
while the expected exclusion range is $\limexplow\GeV \leq \mH \leq \limexphigh\GeV$.

% Acknowledgements for papers with collision data
% Version 9-Feb-2012

\section{Acknowledgements}

We thank CERN for the very successful operation of the LHC, as well as the
support staff from our institutions without whom ATLAS could not be
operated efficiently.

We acknowledge the support of ANPCyT, Argentina; YerPhI, Armenia; ARC,
Australia; BMWF, Austria; ANAS, Azerbaijan; SSTC, Belarus; CNPq and FAPESP,
Brazil; NSERC, NRC and CFI, Canada; CERN; CONICYT, Chile; CAS, MOST and NSFC,
China; COLCIENCIAS, Colombia; MSMT CR, MPO CR and VSC CR, Czech Republic;
DNRF, DNSRC and Lundbeck Foundation, Denmark; EPLANET and ERC, European Union;
IN2P3-CNRS, CEA-DSM/IRFU, France; GNAS, Georgia; BMBF, DFG, HGF, MPG and AvH
Foundation, Germany; GSRT, Greece; ISF, MINERVA, GIF, DIP and Benoziyo Center,
Israel; INFN, Italy; MEXT and JSPS, Japan; CNRST, Morocco; FOM and NWO,
Netherlands; RCN, Norway; MNiSW, Poland; GRICES and FCT, Portugal; MERYS
(MECTS), Romania; MES of Russia and ROSATOM, Russian Federation; JINR; MSTD,
Serbia; MSSR, Slovakia; ARRS and MVZT, Slovenia; DST/NRF, South Africa;
MICINN, Spain; SRC and Wallenberg Foundation, Sweden; SER, SNSF and Cantons of
Bern and Geneva, Switzerland; NSC, Taiwan; TAEK, Turkey; STFC, the Royal
Society and Leverhulme Trust, United Kingdom; DOE and NSF, United States of
America.

The crucial computing support from all WLCG partners is acknowledged
gratefully, in particular from CERN and the ATLAS Tier-1 facilities at
TRIUMF (Canada), NDGF (Denmark, Norway, Sweden), CC-IN2P3 (France),
KIT/GridKA (Germany), INFN-CNAF (Italy), NL-T1 (Netherlands), PIC (Spain),
ASGC (Taiwan), RAL (UK) and BNL (USA) and in the Tier-2 facilities
worldwide.

\bibliographystyle{atlasnote}
\bibliography{HWWlvlvPaper}

\clearpage
\onecolumn
% ATLAS Collaboration author list for 03-APR-2012
% Data extracted on 06-May-2012 for paperid 220
%\documentclass[11pt]{article}
%\usepackage{a4wide}\begin{document}
\begin{flushleft}
{\Large The ATLAS Collaboration}

\bigskip

G.~Aad$^{\rm 48}$,
B.~Abbott$^{\rm 111}$,
J.~Abdallah$^{\rm 11}$,
S.~Abdel~Khalek$^{\rm 115}$,
A.A.~Abdelalim$^{\rm 49}$,
O.~Abdinov$^{\rm 10}$,
B.~Abi$^{\rm 112}$,
M.~Abolins$^{\rm 88}$,
O.S.~AbouZeid$^{\rm 158}$,
H.~Abramowicz$^{\rm 153}$
% ...
\bigskip

% ...
$^{10}$ Institute of Physics, Azerbaijan Academy of Sciences, Baku, Azerbaijan\\
$^{11}$ Institut de F\'isica d'Altes Energies and Departament de F\'isica de la Universitat Aut\`onoma  de Barcelona and ICREA, Barcelona, Spain\\
% ...
$^{48}$ Fakult\"{a}t f\"{u}r Mathematik und Physik, Albert-Ludwigs-Universit\"{a}t, Freiburg i.Br., Germany\\
$^{49}$ Section de Physique, Universit\'e de Gen\`eve, Geneva, Switzerland\\
% ...
$^{88}$ Department of Physics and Astronomy, Michigan State University, East Lansing MI, United States of America\\
% ...
$^{111}$ Homer L. Dodge Department of Physics and Astronomy, University of Oklahoma, Norman OK, United States of America\\
$^{112}$ Department of Physics, Oklahoma State University, Stillwater OK, United States of America\\
% ...
$^{115}$ LAL, Universit\'e Paris-Sud and CNRS/IN2P3, Orsay, France\\
% ...
$^{153}$ Raymond and Beverly Sackler School of Physics and Astronomy, Tel Aviv University, Tel Aviv, Israel\\
% ...
$^{158}$ Department of Physics, University of Toronto, Toronto ON, Canada\\
% ...
\end{flushleft}

%\end{document}

\end{document}
\bye